# Optical activity and phase transformations in γ/β Ga$_2$O$_3$ bilayers under annealing

Alexander Azarov[1,a], Augustinas Galeckas[1], Ildikó Cora[2], Zsolt Fogarassy[2], Vishnukanthan Venkatachalapathy[1], Eduard Monakhov[1], and Andrej Kuznetsov[1,b]

[1] *University of Oslo, Department of Physics, Centre for Materials Science and Nanotechnology, PO Box 1048 Blindern, N-0316 Oslo, Norway*

[2] *HUN-REN Centre for Energy Research, Institute of Technical Physics and Materials Science, 1121 Budapest, Konkoly-Thege M. u. 29-33, Hungary*

**Abstract**

Gallium oxide (Ga$_2$O$_3$) can be crystallized in several polymorphs exhibiting different physical properties. In this work, polymorphic structures consisting of the cubic defective spinel (γ) film on the top of the monoclinic (β) substrate were fabricated by disorder-induced ordering, known to be a practical way to stack these polymorphs together. Such bilayer structures were annealed to investigate the optical properties and phase transformations. Specifically, photoluminescence and diffuse reflectance spectroscopies were combined with transmission electron microscopy, Rutherford backscattering/channeling spectrometry and x-ray diffraction to monitor the evolutions. As a result we observe a two-stage annealing kinetics in γ/β Ga$_2$O$_3$ bilayers associated with the epitaxial γ-to-β regrowth at the interface at temperatures below 700 °C and a non-planar γ-to-β phase transformation starting at higher temperatures. Thus, the present data enhance understanding of the polymorphism in Ga$_2$O$_3$, interconnecting the phase transformation kinetics with the evolution of the optical properties.

## 1 Introduction

Ultra-wide bandgap semiconductors, such as gallium oxide (Ga$_2$O$_3$), have attracted unprecedented research interest during the past decade [1]. Such attention is stimulated by their unique material properties offering numerous potential applications, e.g. in power electronics and optoelectronic devices capable of operating in the deep UV spectral range [2-5]. Equally importantly, Ga$_2$O$_3$ can be crystallized in several polymorphs having different structures [6-9] and, consequently, physical properties. The monoclinic (β) Ga$_2$O$_3$ phase is thermodynamically stable at normal conditions, while the corundum analogue rhombohedral (α), defect spinel analogue cubic (γ), orthorhombic (κ), and bixbyite analogue cubic (δ) phases are metastable [6-9]. As such polymorphism of Ga$_2$O$_3$ may provide additional functionalization opportunities [10]; for example, if fabricating bilayer consisting of different polymorphs.

According to thermodynamics, polymorphic transformation is a function of temperature and pressure. However, recently it was demonstrated that β-to-γ phase transitions in Ga$_2$O$_3$ can be governed by the disorder-induced ordering during ion bombardment [11-13]. Spectacularly, such transition occurs after reaching the disorder threshold [14] resulting in the formation of the γ/β double layer structure having an abrupt interface. Notably, such structures exhibited remarkably high radiation tolerance, limited primarily by the stoichiometric lattice distortions due to incorporation of the implanted ions [13]. The perspective of using γ-$Ga_2O_3$ for radiation tolerant devices was recently tested in the corresponding Schottky diodes [15].

The present study aims to investigate optical properties in such γ/β $Ga_2O_3$ bilayers as a function of the annealing temperature, in correlation with the phase transformation monitoring. It should be noted that by far there are only a few initial studies of the thermal stability in γ-$Ga_2O_3$ polymorph [16-18]. Here we demonstrate a two-stage annealing kinetics in γ/β $Ga_2O_3$ bilayers associated with the epitaxial γ-to-β regrowth at the interface at temperatures below 700 °C and a non-planar γ-to-β phase transformation starting at higher temperatures. Thus, the present data enhance understanding the polymorphism in $Ga_2O_3$, interconnecting the phase transformation kinetics with the evolution of the functional properties.

## 2 Results and Discussion

The evolution of the γ/β double layer structure in the course of thermal annealing is illustrated by Figs. 1(a) and (b) showing the RBS/C spectra of the structure along with corresponding XRD 2Θ scans. For the as-fabricated state of the γ/β bilayer, the RBS/C spectrum exhibits a characteristic box-like shape "inclined" towards the surface, known as a fingerprint of the crystalline γ-$Ga_2O_3$ film [13]. The bilayer state of the initial sample is further confirmed by the XRD data showing both the (020) β-$Ga_2O_3$ reflection and the (440) γ-$Ga_2O_3$ peak centered at ~63.75° in good agreement with literature value of 63.85° [20]. Notably, the as-fabricated γ film is ~300 nm thick. However, this thickness decreases in the course of anneals as clearly seen from the gradual narrowing of the RBS/C box-like signature. For example, upon 500 °C annealing step its thickness is down to ~275 nm, see Fig. 1(a). This reduction of the γ layer thickness maybe interpreted as γ-to-β polymorph regrowth at the interface towards the surface. At the same time, the γ-phase XRD peak moves closer to the tabulated position of the (440) γ-$Ga_2O_3$ planes indicating structural improvements, see Fig. 1(b).

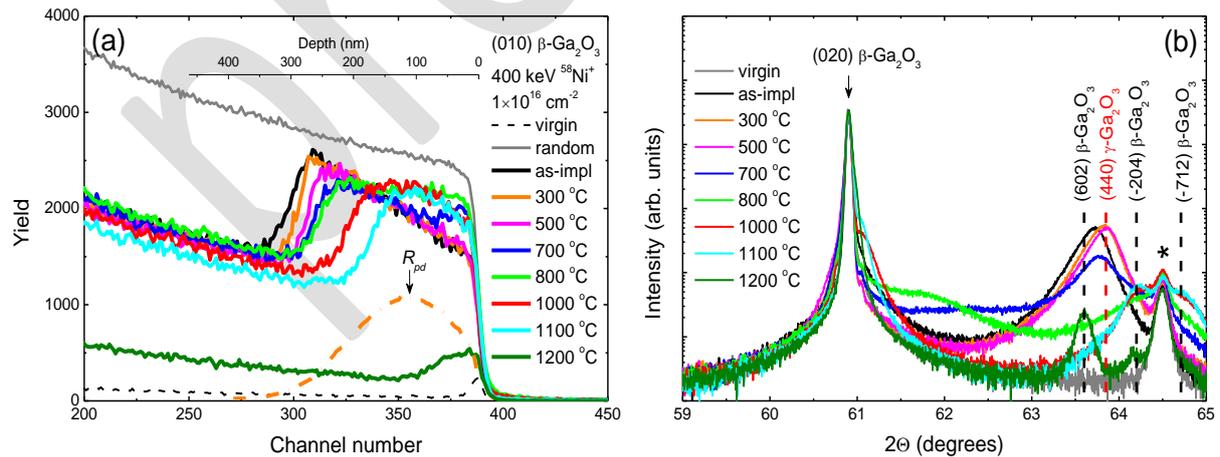

**Figure 1.** Systematic structural analysis of γ/β bilayer anneals. (a) RBS/C spectra and (b) corresponding XRD 2Θ scans of γ/β $Ga_2O_3$ bilayers before and after annealing as indicated in the legends. The peak marked by star in the panel (b) is because of the sample holder. The RBS/C and XRD spectra of the unimplanted (virgin) β-$Ga_2O_3$ samples are also included in the corresponding panels for comparison. Dashed-dotted line illustrates the nuclear energy loss vs depth profile calculated with SRIM code [19], with arrow point the $R_{pd}$.

However, for anneals at ≥700 °C this trend is interrupted and new features are observed in Fig. 1 Indeed, as can be clearly seen in Fig. 1(a), the channeling yield of the 700 °C annealed sample is higher near the surface as compared to that of the samples annealed at lower temperatures. This behavior correlates with a significant decrease in the amplitude of the γ-related XRD peak also accompanied with enhancement of the background intensity between the (020) β- and (440) γ-$Ga_2O_3$ reflections. Further, upon 800 °C anneal the (440) γ-$Ga_2O_3$ peak vanishes as seen from Fig. 1(b). Instead, the XRD peaks centered at ~64.2° and ~64.7° emerge; which can be assigned to the (-204) and (-712) reflections in β-$Ga_2O_3$, respectively [21], indicating misoriented reconstruction relative to the wafer orientation. Thus, this new trend may be interpreted as γ-to-β transformation resulting in differently oriented β-phase in the vicinity of the surface in accordance with the RBS/C data. Indeed, anneals at 1000 and 1100 °C, lead towards even better definition of the XRD peaks; however, the corresponding RBS/C data confirm high dechanneling with the hypothesis of the β-phase misorientation. In contrast, upon 1200 °C anneal, the channeling is considerably improved, even though it still deviates from the virgin signal near the surface. The XRD 2Θ scan measured after 1200 °C anneal confirms further crystalline quality improvement; however, also detecting a new peak centered at 63.6° corresponding to the (602) β-$Ga_2O_3$ reflection [21], in addition to the peaks at 64.2° and 64.7° (Fig. 1(b)).

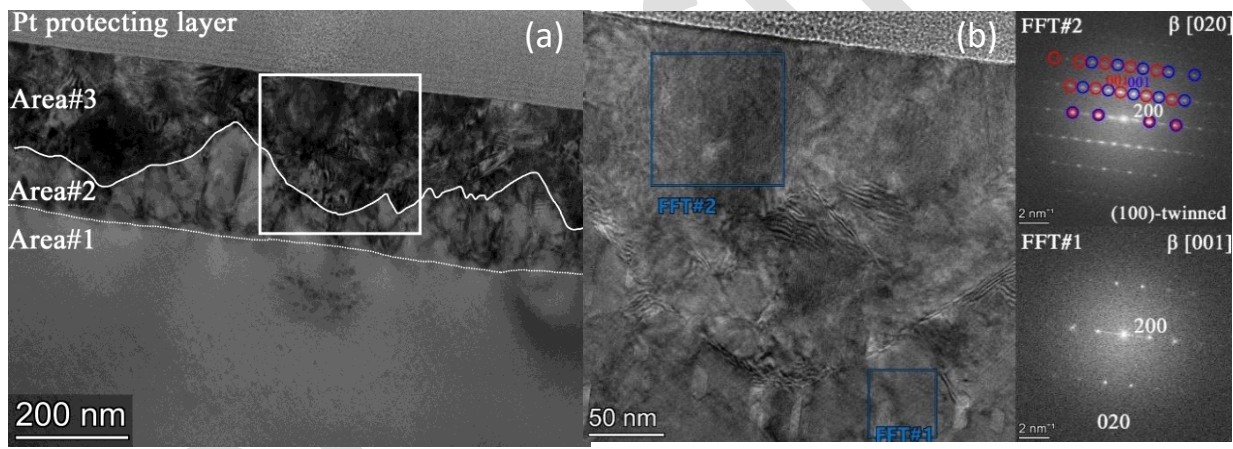

**Figure 2.** Detailed electron microscopy analysis of the sample annealed at 1100 °C (a) Bright field TEM image of the sample. (b) HRTEM image of the area selected by white-colored frame in panel (a) together with the FFTs taken from the two different places in the sample as indicated by blue-colored frames in the image. Area#2 has the same orientation as the underlying (virgin) β substrate (i.e. Area#1), while the reconstructed textured (twinned) β-phase is observed in the Area#3.

To verify the trends observed in Fig. 1, the sample annealed at 1100 °C was selected for detailed TEM study as summarized in Fig. 2. Indeed, the bright field TEM image in Fig. 2(a) shows that the sample consist of three distinct layers marked as Areas #1, #2 and #3. Area #1 corresponds to the initial β-$Ga_2O_3$ substrate. In the Area #2, β-phase was epitaxial regrown keeping the initial substrate orientation. In contrast, the near-to-surface region labeled as Area #3 keeps [200] "in plane" orientation (i.e., perpendicular to the original [020] direction). Furthermore, in this region the β-phase shows (100) twinning as clearly shown by the FFTs (Figs. 2(c)-(d)) taken from the corresponding areas marked in the high resolution TEM image in Fig. 2(b). In addition, it should be mentioned that the interface between Areas #2 and #3 is far from being abrupt. Thus, despite the fact that upon 1100 °C anneal the modified layer has been fully transformed back to the β-phase, see Fig.

2, the near-to-surface misoriented layer gives rise to a higher RBS/C yield as compared to the lower temperatures, see Fig. 1(a). Notably, these trends are not fully consistent with those observed during *in-situ* anneals of similar γ/β structure performed in vacuum [18], detecting the γ/β mixture already after 500 °C anneal; concurrently the β-phase misorientation was observed in the in-situ anneals too [18]. The discrepancies in trends observed in our experiment and during in-situ anneals may be attributed to the factors related to the annealing ambient and potentially different strain effects in macroscopic and TEM-lamella-sized samples.

At this end, under conditions of our experiment, we observe a distinct two-stage annealing kinetics in γ/β bilayered structures. The first stage occurs at temperatures <700°C and is characterized by the shrinkage of the γ layer accompanied with the improvement of the γ-phase crystallinity. The second stage occurs at higher temperatures and is accompanied with the formation of the misoriented β-phase layer in the near-surface region. In the rest of the paper we study how this two-stage structural transformation correlates with optical properties.

Figure 3 summarizes optical absorption properties deduced from DRS measurements carried out at RT. Firstly, Fig. 3(a) and (b) represent selected raw data in semi-logarithmical and normalized scales, respectively. Two characteristic absorption onsets are observed in spectra of these samples, which is likely a signature of dissimilar optical band-edges for different light polarizations [23]. For clarity, in the Tauc plot analysis we consider the lowest onset as a representative of the optical bandgap, see Fig. 3(b). This procedure was systematically applied to all samples and Fig. 3(c) summarizes the data. The trend observed in Fig. 3(c) suggests there is a threshold temperature range – indicated by a highlighted region in Fig. 3(c) - beyond which γ/β double layer structure is apparently converted into a single phase β-$Ga_2O_3$. This agrees well with a similar conclusion reached upon the analysis of the structural transformation trends in Figs. 1-2.

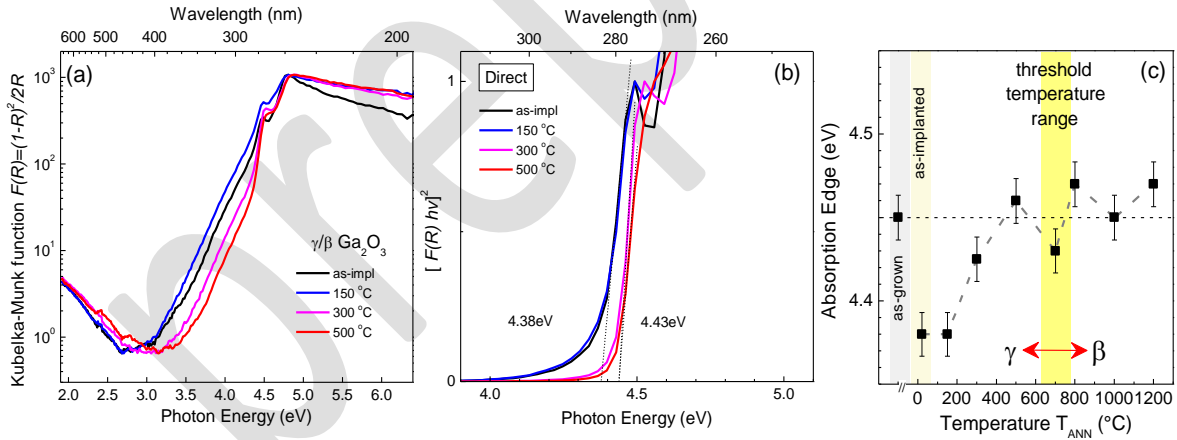

**Figure 3**. Evolution of the optical absorption in γ/β $Ga_2O_3$ bilayers as a function of annealing temperature, deduced from the diffuse-reflectance spectroscopy (DRS) at 300K. (a) Absorbance spectra for selected samples represented by Kubelka-Munk function *F(R)*. (b) Corresponding Tauc plot for indirect optical transitions; straight lines intersecting with the photon energy axis define the absorption-edge positions. (c) summarizing of the optical bandgap (absorption-edge) data.

Optical emission properties are summarized in Fig. 4. PL spectroscopy is known to be extremely sensitive to crystal quality modifications caused by irradiation and post-irradiation [24]. Indeed, as seen from Fig. 4 both the total quantum efficiency (QE) and the spectral response were affected in our samples. This is noticeable already from the raw data in Fig. 4(a), but is even more apparent in Fig. 4(b), where spectra are shown after normalization with regard to the characteristic for $Ga_2O_3$, so-called self-trapped hole (STH) emission centered at ~3.2 eV. One can observe dramatic changes in the green luminescence (GL) at around 2.5 eV and red luminescence (RL) at around 2 eV regions, respectively. In order to make more quantitative estimates, the results of the differential PL analysis are represented by corresponding shaded areas in Fig. 4(b). The annealing leads to a gradual recovery of the crystallinity, as can be seen from the evolution of the total QE in Fig. 4(c), also activating the defect-related RL band, which steadily intensifies up to 800 °C and then rapidly turns off at higher temperatures. Such RL-band evolution may be correlated with the intrinsic properties of the defective spinel structure as well as its transition in to β-phase in agreement with the data in Figs. 1 and 2.

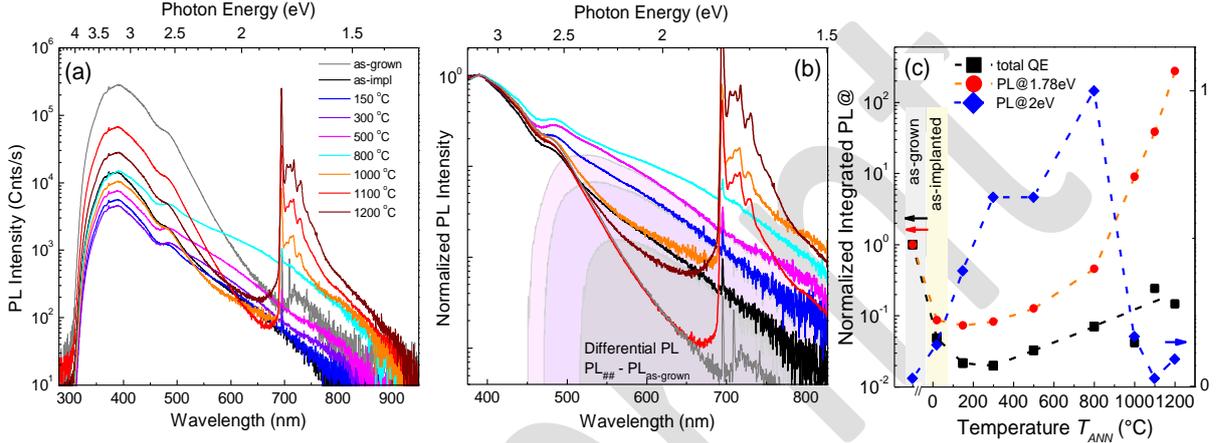

**Figure 4**. Evolution of the optical emission in γ/β $Ga_2O_3$ bilayers as a function of annealing temperature, deduced from photoluminescence (PL) spectroscopy at 10K. (a) PL spectra as a function of annealing temperature. (b) PL spectra normalized with regard to the intrinsic (excitonic) STH emission centered at 3.2eV emphasizing the emission range associated with the implantation-induced defects. Shaded areas represent differential PL signatures obtained by subtracting emission of the unirradiated sample. (c) Evolution of total quantum efficiency (total QE), $Cr^{3+}$ (R1/R2@1.78eV) and defect-related emission (RL(@2eV) as a function of annealing temperature. Note that total QE and $Cr^{3+}$ evolutions are plotted on a semi-logarithmic scale (left-hand side axis), whereas defect-related intensities are on a linear scale (right-hand side axis).

For completeness of the PL data description, we admit that there is another prominent emission component centered at ~694 nm (1.78 eV); as discussed in Supporting Information, this emission is related to Cr [26]. This spectral feature is because of Cr contamination artifact; however, prominent enough not to be ignored assince it may strongly affect the optical performance.

## 3 Conclusions

In conclusion, we investigated the impacts of annealing on the phase transformations and corresponding optical response in γ/β $Ga_2O_3$ double layer structures initially fabricated by disorder-induced ordering. The results show that annealing kinetics of the γ/β $Ga_2O_3$ bilayer exhibits a distinct two-stage behavior associated with the epitaxial γ-to-β recrystallization at the interface at temperatures below 700 °C and a non-planar γ-to-β phase transformation at higher temperatures. Furthermore, non-planar transformations is accompanied with the β-phase misoriention inclusions in the near surface layer from that in the substrate. The optical properties are consistent with the structural transformations, both in terms of γ-to-β bandgap switching and the evolution of the intrinsic defect emissions.

# 4. Experimental Section

In the present work, we used commercial (010) oriented β-Ga$_2$O$_3$ single crystals obtained from Tamura Corp. The γ/β Ga$_2$O$_3$ double layer structures were formed by means of room temperature (RT) bombardment with 400 keV $^{58}$Ni$^+$ ions to a dose of $1\times10^{16}$ cm$^{-2}$. The ion implantation was performed maintaining 7° off-angle orientation from normal direction to minimize channeling and the ion flux was kept constant at $5\times10^{12}$ at/(cm$^2$s). This process is known to result in the γ/β double layer with abrupt interface [13]. Consequently, such γ/β bilayers were isochronically (30 min) annealed in the range of 300 - 1200 °C in air using a conventional tube furnace.

All samples were analyzed by a combination of the Rutherford backscattering spectrometry in channeling mode (RBS/C), x-ray diffraction (XRD), photoluminescence (PL) and diffuse-reflectance spectroscopy (DRS). In addition, selected samples were investigated by transmission electron microscopy (TEM). The RBS/C measurements were performed by 1.6 MeV He$^+$ ions incident along [010] direction in β-Ga$_2$O$_3$ part of the structure and backscattered into a detector placed at 165° relative to the incident beam direction. XRD 2Θ measurements were performed using the Bruker AXS D8 Discover diffractometer with high-resolution Cu K$_{α1}$ radiation.

The highly contrasting thicknesses of γ and β phases in double layer structures (c.f. 300 nm vs 300 μm, i.e. thickness ratio ~1:1000) precludes the use of the most common approach in studying absorption, transmittance spectroscopy, where the signal is integrated over entire double layer structure making contribution from the relatively thin γ-phase layer indistinguishable from the dominating β-phase background. As an alternative, we employ diffuse-reflectance spectroscopy (DRS) technique, which by very definition has enhanced sensitivity to near-to-surface material properties (due to evanescent light penetration, surface-roughness scattering, contrasting refraction in double layer structure, etc.). DRS measurements were carried out at RT using a UV-Vis spectrophotometer (Thermo Scientific, EVO600). The absorption spectra of the samples are represented by Kubelka-Munk function *F(R)*, which is equivalent of absorbance [22].

PL spectra generally comprise emission components describing the intrinsic material properties and those introduced by ion implantation and post-annealing. In Ga$_2$O$_3$ the former is represented by exciton related band, self-trapped holes (STH) to be exact [25]. The ion implantation induced defects can provide non-radiative recombination pathways thus decreasing total QE, and may also introduce luminescent centers that can be observed in PL spectra. The differential PL approach is used to visualize the implantation-induced defect contribution to emission spectra. This is done by subtracting spectrum of the non-implanted sample from all emission spectra. The photo-excitation at 246 nm wavelength (5.04 eV) and 10 mW average power was provided by a third-harmonic of the pulsed Ti:sapphire laser operating at 80 MHz in femtosecond mode-locked regime (Spectra-Physics, Tsunami HP and GWU-UHG-23). PL emission was collected by a microscope and analyzed by a fiber-optic spectrometer (Avantes, AvaSpec-Mini3648-UVI25) covering the wavelength range 200-1100 nm.

The transmission electron microscopy (TEM) investigations were carried out in an aberration-corrected THEMIS microscope (THEMIS 200 TEM/STEM, Thermofisher Scientific/FEI) at 200 keV. The EDS maps were recorded in STEM mode using a Super-X detector. Preparation of the cross-sectional TEM lamella was done in a SCIOS-2 type dual-beam FEG-SEM/FIB. TEM images and diffraction patterns were recorded at 200 keV with a 4k×4k CETA 16 CMOS camera. High resolution TEM (HRTEM) images and corresponding fast Fourier transforms (FFT) were recordered from selected parts of samples too.

## Supporting Information

Supporting Information is available from the Wiley Online Library.

## Acknowledgments

M-ERA.NET Program is acknowledged for financial support via GOFIB project (administrated by the Research Council of Norway project number 337627). AA and EM acknowledge the Research Centre for Sustainable Solar Cell Technology (FME SuSolTech, RCN project number 257639). The Research Council of Norway is also acknowledged for the support to the Norwegian Micro- and Nano-Fabrication Facility, NorFab, project number 295864. IC and ZF acknowledge funding from the Hungarian national project TKP2021-NKTA-05, VEKOP-2.3.3-15-2016-00002 of the European Structural and Investment Funds and János Bolyai Research Scholarship of the Hungarian Academy of Sciences. The authors also gratefully thank Noémi Szász for TEM sample preparations.

## Conflict of interest

The authors have no conflicts to disclose.

## Data availability

The data that support the findings of this study are available from the corresponding author upon reasonable request.

## References:

1. J. Y. Tsao, S. Chowdhury, M. A. Hollis, D. Jena, N. M. Johnson, K. A. Jones, R. J. Kaplar, S. Rajan, C. G. Van de Walle, E. Bellotti, C. L. Chua, R. Collazo, M. E. Coltrin, J. A. Cooper, K. R. Evans, S. Graham, T. A. Grotjohn, E. R. Heller, M. Higashiwaki, M. S. Islam, P. W. Juodawlkis, M. A. Khan, A. D. Koehler, J. H. Leach, U. K. Mishra, R. J. Nemanich, R. C. N. Pilawa-Podgurski, J. B. Shealy, Z. Sitar, M. J. Tadjer, A. F. Witulski, M. Wraback, and J. A. Simmons, "Ultrawide-bandgap semiconductors: Research opportunities and challenges", Adv. Electron. Mater. **4**, 1600501 (2018).

2. S. J. Pearton, J. Yang, P. H. Cary, F. Ren, J. Kim, M. J. Tadjer, and M. A. Mastro, "A review of $Ga_2O_3$ materials, processing, and devices", Appl. Phys. Rev. **5**, 011301 (2018).

3. M. J. Tadjer, "Toward gallium oxide power electronics", Science **378**, 724 (2022).

4. X. Chen, F. Ren, S. Gu, and J. Ye, "Review of gallium-oxide-based solar-blind ultraviolet photodetectors", Photonics Research **7**, 381-415 (2019).

5. R. Zhu, H. Liang, S. Liu, Y. Yuan, X. Wang, F. C.-C. Ling, A. Kuznetsov, G. Zhang, and Z. Mei, "Non-volatile optoelectronic memory based on a photosensitive dielectric", Nat. Commun. **14**, 5396 (2023).

6. H. Y. Playford, A. C. Hannon, E. R. Barney, and R. I. Walton, "Structures of uncharacterised polymorphs of gallium oxide from total neutron diffraction", Chem.-A Eur. J. **19**, 2803 (2013).

7. I. Cora, F. Mezzadri, F. Boschi, M. Bosi, M. Caplovicova, G. Calestani, I. Dodony, B. Pecz, and R. Fornari, "The real structure of ε-$Ga_2O_3$ and its relation to κ-phase", Cryst. Eng. Comm. **19**, 1509 (2017).

8. T. Kato, H. Nishinaka, K. Shimazoe, K. Kanegae, and M. Yoshimoto, "Demonstration of bixbyite-structured δ-$Ga_2O_3$ thin films using β-$Fe_2O_3$ buffer layers by mist chemical vapor deposition", ACS Appl. Electron. Mater. **5**, 1715 (2023).

9. L. Li, W. Wei, and M. Behrens, "Synthesis and characterization of α-, β-, and γ-$Ga_2O_3$ prepared from aqueous solutions by controlled precipitation", Solid State Sci. **14,** 971 (2012).

10. D. Gentili, M. Gazzano, M. Melucci, D. Jones, and M. Cavallini, "Polymorphism as an additional functionality of materials for technological applications at surfaces and interfaces", Chem. Soc. Rev. **48**, 2502 (2019).

11. T. Yoo, X. Xia, F. Ren, A. Jacobs, M. J. Tadjer, S. Pearton, and H. Kim, "Atomic-scale characterization of structural damage and recovery in Sn ion-implanted β-$Ga_2O_3$", Appl. Phys. Lett. **121**, 072111 (2022).

12. H.-L. Huang, C. Chae, J. M. Johnson, A. Senckowski, S. Sharma, U. Singisetti, M. H. Wong, and J. Hwang, "Atomic scale defect formation and phase transformation in Si implanted β-$Ga_2O_3$", APL Mater. **11**, 061113 (2023).

13. A. Azarov, J. García Fernández, J. Zhao, F. Djurabekova, H. He, R. He, Ø. Prytz, L. Vines, U. Bektas, P. Chekhonin, N. Klingner, G. Hlawacek, and A. Kuznetsov, "Universal radiation tolerant semiconductor", Nat. Commun. **14**, 4855 (2023).

14. J. Zhao, J. García-Fernández, A. Azarov, R. He, Ø. Prytz, K. Nordlund, M. Hua, F. Djurabekova, and A. Kuznetsov, "Crystallization instead amorphization in collision cascades in gallium oxide", https://arxiv.org/abs/2401.07675

15. A. Y. Polyakov, A. A. Vasilev, A. I. Kochkova, I. V. Shchemerov, E. B. Yakimov, A. V. Miakonkikh, A. V. Chernykh, P. B. Lagov, Y. S. Pavlov, A. S. Doroshkevich, R. Sh. Isaev, A. A. Romanov, L. A. Alexanyan, N. Matros, A. Azarov, A. Kuznetsov, and S. Pearton, "Proton damage effects in double polymorph γ/β-$Ga_2O_3$ diodes", J. Mat. Chem. C **12** 1020 (2024).

16. C. Wouters, M. Nofal, P. Mazzolini, J. Zhang, T. Remmele, A. Kwasniewski, O. Bierwagen, and M. Albrecht, "Unraveling the atomic mechanism of the disorder–order phase transition from γ-$Ga_2O_3$ to β-$Ga_2O_3$", APL Mater. **12**, 011110 (2024).

17. S. B. Kjeldby, A. Azarov, P. D. Nguyen, V. Venkatachalapathy, R. Mikšová, A. Macková, A. Kuznetsov, Ø. Prytz, and L. Vines, "Radiation-induced defect accumulation and annealing in Si-implanted gallium oxide," J. Appl. Phys. **131**, 125701 (2022).

18. J. García-Fernández, S. B. Kjeldby, L. Zeng, A. Azarov, A. Pokle, P. D. Nguyen, E. Olsson, L. Vines, A. Kuznetsov, and Ø. Prytz, "*In situ* atomic-resolution study of transformations in double polymorph γ/β $Ga_2O_3$ structures", Materials Advances **5**, 3824 (2024).

19. J. F. Ziegler, M. D. Ziegler, and J. P. Biersack, "SRIM—the stopping and range of ions in matter (2010)", Nucl. Instrum. Meth. B **268,** 1818 (2010).

20. A. F. M. Anhar Uddin Bhuiyan, Z. Feng, J. M. Johnson, H.-L. Huang, J. Sarker, M. Zhu, M. R. Karim, B. Mazumder, J. Hwang, and H. Zhao, "Phase transformation in MOCVD growth of $(Al_xGa_{1−x})_2O_3$ thin films", APL Mater. **8**, 031104 (2020).

21. Powder diffraction file PDF #01-087-1901

22. M. L. Myrick, M. N. Simcock, M. Baranowski, H. Brooke, S. L. Morgan, and J. N. Mc Cutcheon, "The Kubelka-Munk diffuse reflectance formula revisited", Appl. Spectroscopy Rev., **46,** 140 (2011).

23. F. Ricci, F. Boschi, A. Baraldi, A. Filippetti, M. Higashiwaki, A. Kuramata, V. Fiorentini, and R. Fornari, "Theoretical and experimental investigation of optical absorption anisotropy in β-$Ga_2O_3$", J. Phys.: Condens. Matter. **28**, 224005 (2016).

24. A. Azarov, A. Galeckas, E. Wendler, E. Monakhov, and A. Kuznetsov, "Inverse dynamic defect annealing in ZnO", Appl. Phys. Lett. 124, 042106 (2024).

25. Y. Wang, P. T. Dickens, J. B. Varley, X. Ni, E. Lotubai, S. Sprawls, F. Liu, V. Lordi, S. Krishnamoorthy, S. Blair, K. G. Lynn, M. Scarpulla, and B. Sensale-Rodriguez, "Incident wavelength and polarization dependence of spectral shifts in β-$Ga_2O_3$ UV photoluminescence", Sci. Rep. **8**, 18075 (2018).

26. J. E. Stehr, M. Jansson, D. M. Hofmann, J. Kim, S. J. Pearton, Weimin M. Chen, and I. A. Buyanova, "Magneto-optical properties of $Cr^{3+}$ in β-$Ga_2O_3$", Appl. Phys. Lett. **119**, 052101 (2021).